\documentclass[a4paper,10pt,twoside]{cpc-hepnp}
\usepackage{CJK,upgreek,fancyhdr}
\usepackage{multicol}
\usepackage{graphicx}
\usepackage{booktabs}
\usepackage{amssymb,bm,mathrsfs,bbm,amscd}
\usepackage[tbtags]{amsmath}
\usepackage{lastpage}
\usepackage[english]{babel}
\usepackage{epsfig}
\usepackage{graphicx}
\usepackage{epstopdf}

\begin{document}

\fancyhead[c]{\small Chinese Physics C~~~Vol. xx, No. x (2020) xxxxxx}
\fancyfoot[C]{\small 010201-\thepage}
\footnotetext[0]{Received 31 June 2020}

\title{Search for the correlations between host properties and ${\rm DM_{host}}$ of fast radio bursts: constraints on the baryon mass fraction in IGM\thanks{Supported by the National Natural Science Fund of China under grant Nos. 11775038, 11873001 and 12147102.}}

\author{%
Hai-Nan Lin$^{1;1)}$\email{linhn@cqu.edu.cn}
\quad Xin Li$^{1}$
\quad Li Tang$^{2}$
}
\maketitle

\address{%
$^1$ Department of Physics, Chongqing University, Chongqing 401331, China\\
$^2$ Department of Math and Physics, Mianyang Normal University, Mianyang 621000, China
}

\begin{abstract}
  The application of fast radio bursts (FRBs) as probes to investigate astrophysics and cosmology requires the proper modelling of the dispersion measures of Milky Way (${\rm DM_{MW}}$) and host galaxy (${\rm DM_{host}}$). ${\rm DM_{MW}}$ can be estimated using the Milky Way electron models, such as NE2001 model and YMW16 model. However, ${\rm DM_{host}}$ is hard to model due to limited information on the local environment of FRBs. In this paper, using 17 well-localized FRBs, we search for the possible correlations between ${\rm DM_{host}}$ and the properties of host galaxies, such as the redshift, the stellar mass, the star-formation rate, the age of galaxy, the offset of FRB site from galactic center, and the half-light radius. We find no strong correlation between ${\rm DM_{host}}$ and any of the host property. Assuming that ${\rm DM_{host}}$ is a constant for all host galaxies, we constrain the fraction of baryon mass in the intergalactic medium today to be $f_{\rm IGM,0}=0.78_{-0.19}^{+0.15}$. If we model ${\rm DM_{host}}$ as a log-normal distribution, however, we obtain a larger value, $f_{\rm IGM,0}=0.83_{-0.17}^{+0.12}$. Based on the limited number of FRBs, no strong evidence for the redshift evolution of $f_{\rm IGM}$ is found.
\end{abstract}

\begin{keyword}
fast radio bursts  --  intergalactic medium  --  cosmological parameters
\end{keyword}

\begin{pacs}
98.54.Gr, 98.58.-w, 98.80.-k
\end{pacs}

\footnotetext[0]{\hspace*{-3mm}\raisebox{0.3ex}{$\scriptstyle\copyright$}2019
Chinese Physical Society and the Institute of High Energy Physics
of the Chinese Academy of Sciences and the Institute
of Modern Physics of the Chinese Academy of Sciences and IOP Publishing Ltd}%

\begin{multicols}{2}

\section{Introduction}\label{sec:introduction}

Fast radio bursts (FRBs) are short-duration and luminous radio transients happening in the Universe, see e.g. Refs.\cite{Petroff:2019tty,Cordes:2019cmq,Zhang:2020qgp,Xiao:2021omr} for recent review. In 2007, Lorimer et al. \cite{Lorimer:2007qn} reanalyzed the archive data of the Parkes 64-m telescope recorded in 2001, and found an extraordinary radio pulse, which is now named FRB 010724. This phenomenon is for long time not attracted much attention until four other bursts were discovered several years later \cite{Thornton:2013iua}. Thereafter, FRBs have attracted great interests within the astronomy community. The observed dispersion measures (DM) of most FRBs significantly exceed the contribution from the Milky Way, hinting that they occur at cosmological distance. The cosmological origin was further confirmed as the identification of host galaxy and the direct measurement of redshift \cite{Keane:2016yyk,Chatterjee:2017dqg,Tendulkar:2017vuq}. There are in general two kinds of FRBs, i.e., the repeaters and non-repeaters. Most of the repeating sources found by the Canadian Hydrogen Intensity Mapping Experiment (CHIME) telescope only repeated two or three times for each source \cite{Andersen:2019yex}. There is one exception, FRB 121102, from which thousands of bursts have been observed by different telescopes \cite{Spitler:2016dmz,Scholz:2017kwy,Zhang:2018jux,Gourdji:2019lht,Oostrum:2019fsh,Li:2021hpl}. Statistical analysis of FRB 121102 shows that the burst energies and waiting times follow the power-law distribution \cite{Wang:2016lhy,Wang:2017agh,Wang:2019sio}, hinting that the explosion of repeating FRBs may be a self-organized criticality process. Further analysis shows that the bursts of FRB 121102 can be more well fitted by the bent power-law and are scale-invariant \cite{Lin:2019ldn}, implying that there are some similarities between FRBs and soft gamma repeaters (SGRs) \cite{Chang:2017bnb,Sang:2021cjq}. As the FAST telescope extensively increases the bursts from FRB 121102, a bimodal burst energy distribution is found \cite{Li:2021hpl}. Repeating FRBs usually have no regular period, but the CHIME/FRB Collaboration \cite{Amiri:2020gno} found an unexpected long period of 16.35 days with an approximately 4-day active window in FRB 180916.J0158+65. The physical mechanism of FRBs is still under extensive debate. Several theoretical models have been proposed to explain the explosion of FRBs \cite{Totani:2013lia,Loeb:2013wna,Huang:2015peq,Dai:2016qem,Lyutikov:2016ueh,Wang:2016dgs,Zhang:2017zse,Munoz:2019fta,Dai:2020ctu}. The most popular models involve one or two compact objects such as neutron star and magnetar in the central of FRB source. The recently discovered burst FRB200428, which was associated with a Galactic magnetar, strongly supports the magnetar origin of at least some FRBs \cite{Andersen:2020hvz,Bochenek:2020zxn}.

FRBs are very luminous and they are expected to be detectable in the most ideal case up to redshift $z\sim 15$ for sensitive radio telescopes such as the Five-hundred-meter Aperture Spherical radio Telescope \cite{Zhang:2018csb}. Therefore, FRBs could be used as probes to study the high-redshift cosmology. For example, Munoz et al. \cite{Munoz:2016tmg} pointed out that the strongly lensed FRBs can be used to probe the compact dark matter in the Universe. Yu \& Wang \cite{Yu:2017beg} showed that FRBs can be used to measure the cosmic proper distance. Li et al. \cite{Li:2017mek} proposed that the strongly lensed repeating FRBs can tightly constrain the Hubble constant and cosmic curvature. Walters et al. \cite{Walters:2017afr} showed that FRBs can be used to constrain the baryon matter density. Li et al. \cite{Li:2019klc,Li:2020qei} showed that FRBs can be used to constrain the fraction of baryon mass in the intergalactic medium (IGM). Xu \& Zhang  \cite{Xu:2020jic} proposed that FRBs can be used to probe the intergalactic turbulence. Wu et al. \cite{Wu:2020jmx} proposed that FRBs can be used to model-independently measure the Hubble parameter. Qiang et al. \cite{Qiang:2019zrs} showed that FRBs can be used to test the possible cosmic anisotropy. Pagano \& Fronenberg \cite{Pagano:2021zla} pointed out that the highly dispersed FRBs can be used to constrain the epoch of cosmic reionization. Pearson et al. \cite{Pearson:2020wxb} showed that strongly lensed repeating FRBs can be used as the probes to search for gravitational waves. In addition, FRBs can be used as the probes to test the fundamental physics, such as constraining the Lorentz invariance violation, the weak equivalent principle and the photon mass \cite{Wei:2015hwd,Wu:2016brq,Tingay:2016tgf,Bonetti:2016cpo,Wei:2021vvn}.

The applications of FRBs as the probes to investigate the Universe often involve the observation of dispersion measure (DM, see the next section for detail), which depends on the electron distribution alone the line of sight. The total DM of an extragalactic FRB consists of four parts: the Milky Way interstellar medium (ISM), the Milky Way halo, the IGM and the host galaxy. The electron distribution in the Milky Way ISM has been modelled from pulsar observation, for example, the TC93 model \cite{Taylor:1993my}, NE2001 model \cite{Cordes:2002wz}, YMW16 model \cite{Yao_2017msh}, and so on. The DM of Milky Way halo can be reasonably estimated \cite{Macquart:2020lln}. Therefore, the DM contributed by the Milky Way can be subtracted from the total DM. The IGM contribution is proportional to the electron density in IGM, which depends on the density and ionization rate of baryon matter in IGM. It is this part that contains the information of Universe and can be used to investigate the cosmology. The difficulty is that the host galaxy contribution to DM is hard to model. This is because, although there are some observations \cite{2020arXiv200513158C,2021ApJ...917...75M,2020ApJ...903..152H}, we still have poor knowledge on the local environment of majority of FRBs. Several factors may affect the DM of host galaxy, such as the galaxy type, the inclination angle of host galaxy, the mass of host galaxy, the star-formation rate, the offset of FRB site from galactic center, just to name a few. FRBs have been observed in different types of galaxies, and there is no unique way to model the DM of host galaxy. For example, Xu \& Han \cite{Xu:2015wza} modelled the DM of host galaxy by assuming that the host galaxies are similar to the Milky Way or M31. Luo et al. \cite{Luo:2018tiy} assumed that the distribution of DM of host galaxy follows the star-formation rate (SFR). Because there is a lack of direct measurement on the DM of host galaxy, reasonably extracting it from observation is of great importance. Yang \& Zhang \cite{Yang:2016zbm} pointed out that the average DM of host galaxy can be obtained statistically from a large sample of FRBs with redshift measurements. However, this method requires the reconstruction of the first order derivative curves from discrete data points, which will introduce large uncertainty.

Up to now, hundreds of FRBs are reported \cite{Petroff:2016tcr,CHIMEFRB:2021srp}. However, there are only 19 well-localized FRBs (except for the Galactic FRB200428) which have direct identification of host galaxies\footnote{The FRB Host Database, http://frbhosts.org/}. All of the 19 well-localized FRBs have direct measurement of redshift (either spectroscopic redshift or photometric redshift), which falls in the redshift range $z\in(0.0039,0.66)$. The properties of the host galaxies, such as the stellar mass, the age of galaxy, the SFR, the half-light radius, and so on, have been observed in detail by follow-up observations. In this paper, based on the well-localized FRBs, we investigate the DM of host galaxy statistically. The rest parts of this paper are arranged as follows: In Section 2, we search for the possible correlations between ${\rm DM_{host}}$ and the host galaxy properties. In Section 3, assuming a constant value of ${\rm DM_{host}}$, we use the well-localized FRBs to constrain the fraction of baryon mass in IGM. Finally, discussions and conclusions are given in Section 4.

\section{The DM of host galaxy}\label{sec:DM_host}

The propagation of electromagnetic waves in cold plasma leads to the frequency-dependent group velocity of light. Therefore, photons with different energy travelling the same distance cost different time. This plasma effect, although is tiny, can be detectable if cumulated at cosmological distance. The time delay between low- and high- energy photons propagating from a distant source to earth is proportional to a quantity called dispersion measure (DM), which is the integral of electron density along the photon path \cite{Inoue:2003ga}. The plasma effect is negligible for the visible light, but it is important for the radio waves in e.g. FRBs. The DM of an FRB can be obtained from the time-resolved spectrum. The observed DM of an extragalactic FRB consists of three parts: the contributions from Milky Way, intergalactic medium (IGM) and host galaxy \cite{Deng:2013aga,Gao:2014iva},
\begin{equation}\label{eq:DM_obs}
  {\rm DM_{obs}}={\rm DM_{MW}}+{\rm DM_{IGM}}+\frac{{\rm DM_{host}}}{1+z},
\end{equation}
where ${\rm DM_{host}}$ is the DM of host galaxy in the source frame, $z$ is the redshift of host galaxy, and the factor $1+z$ accounts for the cosmic dilation.

The DM of Milky Way can be divided into two components: the contributions from Milky Way interstellar medium (ISM) and Milky Way halo \cite{Macquart:2020lln},
\begin{equation}
  {\rm DM_{MW}}={\rm DM_{MW,ISM}}+{\rm DM_{MW,halo}}.
\end{equation}
Given the sky position of an FRB, ${\rm DM_{MW,ISM}}$ can be well constrained by modelling the electron distribution in the Milky Way ISM from pulsar observations, such as the NE2001 model \cite{Cordes:2002wz} and the YMW16 model \cite{Yao_2017msh}. The Milky Way halo contribution is not well constrained yet, but it is expected to be in the range $50-100~{\rm pc~cm^{-3}}$ \cite{Macquart:2020lln}.

The DM of IGM, assuming that both hydrogen and helium are fully ionized, can be written as \cite{Deng:2013aga,Zhang:2020ass}
\begin{equation}\label{eq:DM_IGM}
  {\rm \overline{DM}_{IGM}}(z)=\frac{21cH_0\Omega_b}{64\pi Gm_p}\int_0^z\frac{f_{\rm IGM}(z)(1+z)}{\sqrt{\Omega_m(1+z)^3+\Omega_\Lambda}}dz,
\end{equation}
where $m_p$ is the proton mass, $f_{\rm IGM}(z)$ is the fraction of baryon mass in IGM, $H_0$ is the Hubble constant, $G$ is the Newtonian gravitational constant, $\Omega_b$ is the normalized baryon matter density, $\Omega_m$ and $\Omega_\Lambda$ are the normalized densities of matter (includes baryon matter and dark matter) and dark energy at present day, respectively. Note that equation (\ref{eq:DM_IGM}) should be interpreted as the mean contribution from IGM. The actual value would deviate from equation (\ref{eq:DM_IGM}) causing by e.g. the fluctuation of baryon matter, the incomplete ionization of hydrogen or helium, etc.

The DM of host galaxy is difficult to model due to the lack of observation on the local environment of FRB sources. However, given that the DM of Milky Way is modeled, and the DM of IGM is predicted by a specific cosmological model, we can invert equation (\ref{eq:DM_obs}) to obtain the DM of host galaxy,
\begin{equation}\label{eq:DM_host}
  {\rm DM_{host}}=(1+z)({\rm DM_{obs}}-{\rm DM_{MW}}-{\rm DM_{IGM}}).
\end{equation}
The uncertainties of ${\rm DM_{host}}$ can be calculated using the standard error propagating formula,
\begin{equation}\label{eq:sigma_host}
   \sigma_{\rm host}=(1+z)\sqrt{(\sigma_{\rm obs}^2+\sigma_{\rm MW}^2+\sigma_{\rm IGM}^2)},
\end{equation}
where the uncertainty on ${\rm DM_{MW}}$ is propagated from the uncertainties on ${\rm DM_{MW,ISM}}$ and ${\rm DM_{MW,halo}}$,
\begin{equation}
  \sigma_{\rm MW}=\sqrt{\sigma_{\rm MW,ISM}^2+\sigma_{\rm MW,halo}^2}.
\end{equation}
Note that the DM contribution from host galaxy also consists of ISM and halo parts. Without other observations, these two parts are completely degenerated. Therefore, we don't distinguish them and treat them as a whole.

So far there are in total 19 extragalactic FRBs that are well localized and host identified\footnote{The FRB Host Database, http://frbhosts.org/}. Among the 19 FRBs, we omit FRB20200120E and FRB20190614D. The repeating burst FRB20200120E is localized to the direction of M81, and its redshift is measured to be $-0.0001$ \cite{Bhardwaj:2021xaa}. This burst is very close to the Milky Way and the peculiar velocity dominates over the Hubble flow, so it is inappropriate to be used to study the cosmology. The non-repeating burst FRB20190614D has a photometric redshift $z\approx 0.6$ \cite{Law:2020cnm}, but there is a lack of detailed observation on the host galaxy. Therefore, we only consider the rest 17 FRBs, whose properties are listed in Table \ref{tab:host}. Among these 17 FRBs, 6 bursts are repeating and 11 bursts are non-repeating. All the FRBs have well-measured sky position (RA, Dec), the observed dispersion measure (${\rm DM_{obs}}$), the spectroscopic redshift ($z$), the stellar mass of host galaxy ($M_{\star}$), the star formation rate (SFR), the mass-weighted age of host galaxy (Age), the offset of FRB site from galactic center (Offset), and the half-light radius of host galaxy ($R_{\rm eff}$). We calculate the DM of the Milky Way ISM using two different models, i.e. the NE2001 model and YMW16 model, and list the results in the fifth and sixth columns of Table \ref{tab:host}, respectively.

\begin{table*}[htbp]
\footnotesize
\caption{\small{The properties of host galaxies of 17 well-localized FRBs.}}
\label{tab:host}
\tabcolsep 2pt 
\begin{tabular*}{1.1\textwidth}{cccccccccccccc}
\hline\hline
FRBs & RA & Dec & ${\rm DM_{obs}}$ & NE2001 & YMW16 & $z$ & $M_{\star}$ & SFR & Age & Offset & $R_{\rm eff}$ & repeat? & References\\
& [ $^{\circ}$ ] & [ $^{\circ}$ ] & [${\rm pc~cm^{-3}}$] & [${\rm pc~cm^{-3}}$] & [${\rm pc~cm^{-3}}$] &   & [$10^9M_{\odot}$] & [$M_{\odot}/yr$] & [Myr] & [kpc] & [kpc] &  &\\\hline
20121102A & $82.99$ & $33.15$ &557.00 &157.60 &287.62 &0.1927 & $0.14\pm0.07$ & $0.15\pm0.04$ & $257.7$ & $0.8\pm0.1$ & $2.05\pm0.11$ &Yes & \cite{Spitler:2016dmz,Chatterjee:2017dqg,Tendulkar:2017vuq,Bassa:2017tke,2020ApJ...903..152H}\\
20180301A & $93.23$ & $4.67$ &536.00 &136.53 &263.16 &0.3305 & $2.30\pm0.60$ & $1.93\pm0.58$ & $607.2$ & $10.8\pm3.0$ & $5.80\pm0.20$ &Yes & \cite{Bhandari:2021pvj}\\
20180916B & $29.50$ & $65.72$ &348.80 &168.73 &319.42 &0.0337 & $2.15\pm0.33$ & $0.06\pm0.02$ & $154.9$ & $5.5\pm0.0$ & $3.57\pm0.36$ &Yes & \cite{2020ApJ...903..152H,Marcote:2020ljw,CHIMEFRB:2020bcn,Tendulkar:2020npy,2021ApJ...917...75M}\\
20180924B & $326.11$ & $-40.90$ &362.16 &41.45 &27.28 &0.3214 & $13.20\pm5.10$ & $0.88\pm0.26$ & $383.4$ & $3.4\pm0.8$ & $2.75\pm0.10$ &No & \cite{2020ApJ...903..152H,2021ApJ...917...75M,Bannister:2019iju,Bhandari:2020oyb,Day:2020yap}\\
20181030A & $158.60$ & $73.76$ &103.50 &40.16 &32.72 &0.0039 & $5.80\pm1.80$ & $0.36\pm0.10$ & $4800.0$ & $\--$ & $2.60\pm0.00$ &Yes & \cite{Bhardwaj:2021hgc}\\
20181112A & $327.35$ & $-52.97$ &589.00 &41.98 &28.65 &0.4755 & $3.98\pm2.02$ & $0.37\pm0.11$ & $572.4$ & $1.7\pm19.2$ & $7.19\pm1.70$ &No & \cite{2020ApJ...903..152H,Bhandari:2020oyb,2019Sci...366..231P}\\
20190102C & $322.42$ & $-79.48$ &364.55 &56.22 &42.70 &0.2913 & $3.39\pm1.02$ & $0.86\pm0.26$ & $55.6$ & $2.3\pm4.2$ & $5.00\pm0.15$ &No & \cite{2020ApJ...903..152H,2021ApJ...917...75M,Bhandari:2020oyb,Day:2020yap,Macquart:2020lln}\\
20190523A & $207.06$ & $72.47$ &760.80 &36.74 &29.75 &0.6600 & $61.20\pm40.10$ & $0.09\pm0.00$ & $685.9$ & $27.2\pm22.6$ & $3.28\pm0.18$ &No & \cite{2020ApJ...903..152H,Ravi:2019alc}\\
20190608B & $334.02$ & $-7.90$ &340.05 &37.81 &26.44 &0.1178 & $11.60\pm2.80$ & $0.69\pm0.21$ & $383.4$ & $6.5\pm0.8$ & $7.37\pm0.07$ &No & \cite{2020ApJ...903..152H,2021ApJ...917...75M,Bhandari:2020oyb,Day:2020yap,Macquart:2020lln,2020arXiv200513158C}\\
20190611B & $320.74$ & $-79.40$ &332.63 &56.60 &43.04 &0.3778 & $0.75\pm0.53$ & $0.27\pm0.08$ & $\--$ & $11.7\pm5.8$ & $2.15\pm0.11$ &No & \cite{2020ApJ...903..152H,Day:2020yap,Macquart:2020lln}\\
20190711A & $329.42$ & $-80.36$ &592.60 &55.37 &42.06 &0.5217 & $0.81\pm0.29$ & $0.42\pm0.12$ & $607.2$ & $3.2\pm2.1$ & $2.94\pm0.17$ &Yes & \cite{2020ApJ...903..152H,2021ApJ...917...75M,Day:2020yap,Macquart:2020lln}\\
20190714A & $183.98$ & $-13.02$ &504.13 &38.00 &30.94 &0.2365 & $14.20\pm5.50$ & $0.65\pm0.20$ & $1593.2$ & $2.7\pm1.8$ & $3.94\pm0.05$ &No & \cite{2020ApJ...903..152H,2021ApJ...917...75M}\\
20191001A & $323.35$ & $-54.75$ &507.90 &44.22 &30.67 &0.2340 & $46.40\pm18.80$ & $8.06\pm2.42$ & $639.7$ & $11.1\pm0.8$ & $5.55\pm0.03$ &No & \cite{2020ApJ...903..152H,2021ApJ...917...75M,Bhandari:2020cde}\\
20191228A & $344.43$ & $-29.59$ &297.50 &33.75 &19.67 &0.2432 & $5.40\pm6.00$ & $0.03\pm0.01$ & $\--$ & $5.7\pm3.3$ & $1.78\pm0.06$ &No & \cite{Bhandari:2021pvj}\\
20200430A & $229.71$ & $12.38$ &380.25 &27.35 &26.33 &0.1608 & $2.10\pm1.10$ & $0.26\pm0.08$ & $689.5$ & $1.7\pm2.2$ & $1.64\pm0.53$ &No & \cite{2020ApJ...903..152H,Bhandari:2021pvj}\\
20200906A & $53.50$ & $-14.08$ &577.80 &36.19 &38.37 &0.3688 & $13.30\pm3.70$ & $0.48\pm0.14$ & $1150.7$ & $5.9\pm2.0$ & $7.58\pm0.06$ &No & \cite{Bhandari:2021pvj}\\
20201124A & $77.01$ & $26.06$ &413.52 &126.49 &204.74 &0.0979 & $16.00\pm1.00$ & $2.12\pm0.49$ & $5000.0$ & $1.3\pm0.1$ & $\--$ &Yes & \cite{Fong:2021xxj,Ravi:2021kqk,Piro:2021upe}\\
\hline
\end{tabular*}
\end{table*}

\begin{figure*}[htbp]
\centering
\includegraphics[width=1.0\textwidth]{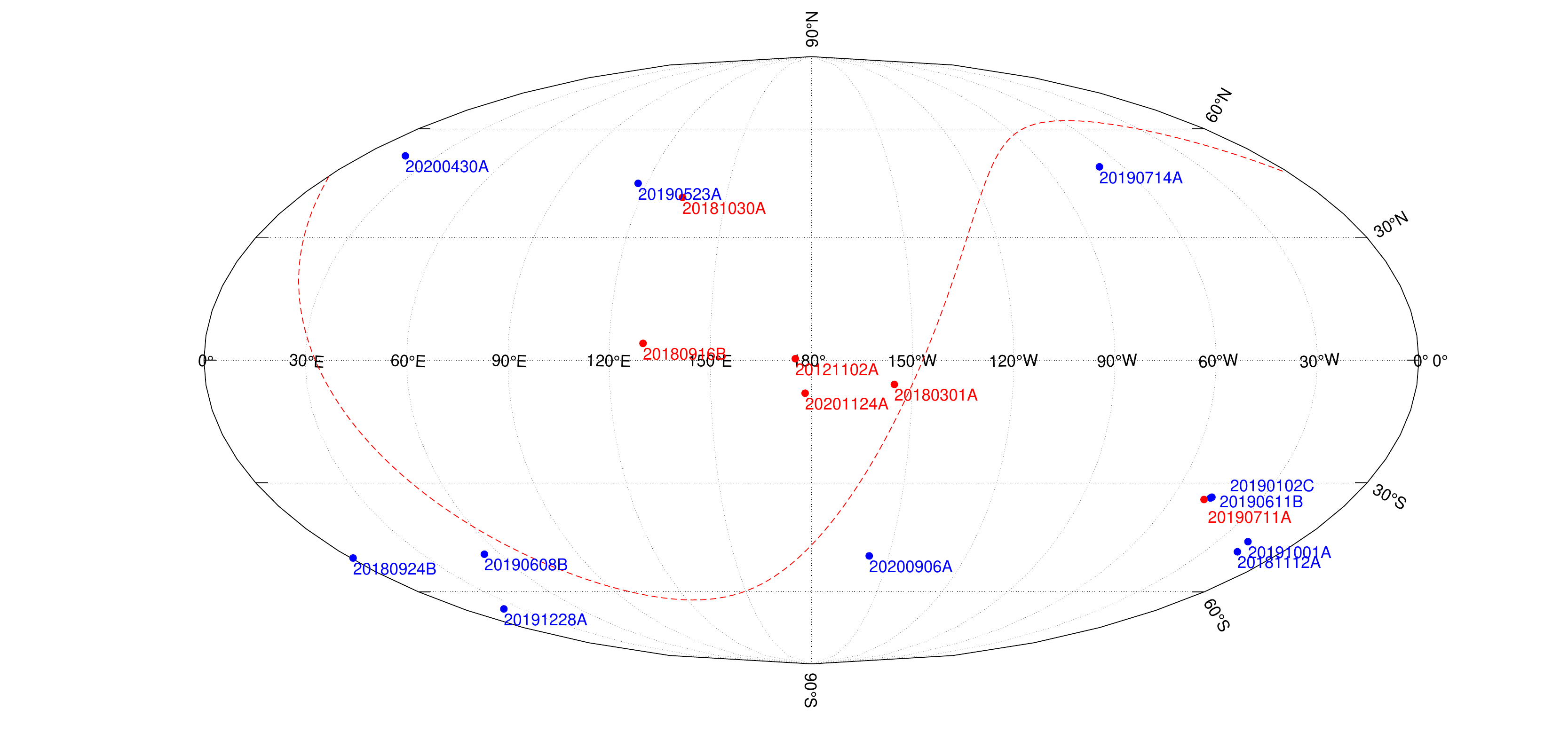}
\caption{\small{The sky positions of 17 well-localized FRBs in the Galactic coordinates. The repeaters and non-repeaters are denoted in red and blue dots, respectively. The red-dashed line is the Equatorial plane.}}
\label{fig:skyposition}
\end{figure*}

Figure \ref{fig:skyposition} shows the sky positions of 17 FRBs in the Galactic coordinates. The repeaters and non-repeaters are denoted in red and blue dots, respectively. Four repeaters (FRB20121102A, FRB20180301A, FRB20180916B and FRB20201124A) locate at low Galactic latitudes, so the Milky Way ISM contribution to DM is very large (see Table \ref{tab:host}). The rest 13 bursts locate at high Galactic latitudes ($|b|>30^{\circ}$), hence the Milky Way ISM contribution to DM is relatively small, with the mean values ${\rm \overline{DM}_{MW,ISM}}=42$ and 32 ${\rm pc~cm^{-3}}$ for NE2001 model and YMW16 model, respectively. Three bursts (FRB20190102C, FRB20190611B and FRB20190711A) have very close sky orientation, hence their Milky Way ISM contributions to DM are similar with each other. We note that ${\rm DM_{MW,ISM}}$ strongly depends on the Milky Way electron models, especially for low-latitude FRBs. At low Galactic latitude, the YMW16 model predicts a much larger value of ${\rm DM_{MW,ISM}}$ than the NE2001 model. At high Galactic latitude, on the contrary, the YMW16 model in general gives a smaller value of ${\rm DM_{MW,ISM}}$ than the NE2001 model.

We calculate the DM of host galaxy, ${\rm DM_{host}}$, by subtracting ${\rm DM_{MW}}$ (including Milky Way ISM and halo contributions) and ${\rm DM_{IGM}}$ from the observed ${\rm DM_{obs}}$ according to equation (\ref{eq:DM_host}). The ${\rm DM_{IGM}}$ term is calculate according to equation (\ref{eq:DM_IGM}) using the Planck 2018 parameters, $H_0=67.4~{\rm km~s^{-1}~Mpc^{-1}}$, $\Omega_m=0.315$, $\Omega_\Lambda=0.685$ and $\Omega_{b0}=0.0493$ \cite{Aghanim:2018eyx}. The fraction of baryon mass is assumed to be a constant, $f_{\rm IGM}=0.84$ \cite{Li:2020qei,Zhang:2020ass}. The uncertainty of ${\rm DM_{host}}$ is calculated using equation (\ref{eq:sigma_host}). The ${\rm DM_{obs}}$ can be tightly constrained by observing the time-resolved spectra of FRBs. According to the FRB catalog \cite{Petroff:2016tcr}, the average uncertainty on ${\rm DM_{obs}}$ is only $\sim 1.5$ pc cm$^{-3}$. Both the NE2001 model and YMW16 model do not provide the uncertainty on ${\rm DM_{MW,ISM}}$. Since these two models predict different values of ${\rm DM_{MW,ISM}}$, we take $\sigma_{\rm MW,ISM}$ as the absolute value of the difference of ${\rm DM_{MW,ISM}}$ calculated using these two models. This ensures that two models give consistent results within $1\sigma$ uncertainty. For FRBs at high Galactic latitude $(|b|>10^\circ)$, the value of $\sigma_{\rm MW,ISM}$ is about 10 pc cm$^{-3}$, while for low-latitude $(|b|<10^\circ)$ FRBs it is at the order of magnitude 100 pc cm$^{-3}$. The Milky Way halo contribution is assumed to be ${\rm DM_{MW,halo}}=50~{\rm pc~cm^{-3}}$ \cite{Macquart:2020lln}, and we add 50\% uncertainty on it. The ${\rm DM_{IGM}}$ term has large uncertainty due to the density fluctuation in the large-scale structure \cite{Pol:2019cfk}. Cosmological simulations show that the uncertainty on ${\rm DM_{IGM}}$ increases with redshift \cite{McQuinn:2013tmc}. Here we use the $\sigma_{{\rm IGM}}(z)$ relation given in Ref.\cite{Li:2019klc} to calculate the uncertainty.

\begin{figure*}[htbp]
\centering
 \includegraphics[width=0.48\textwidth]{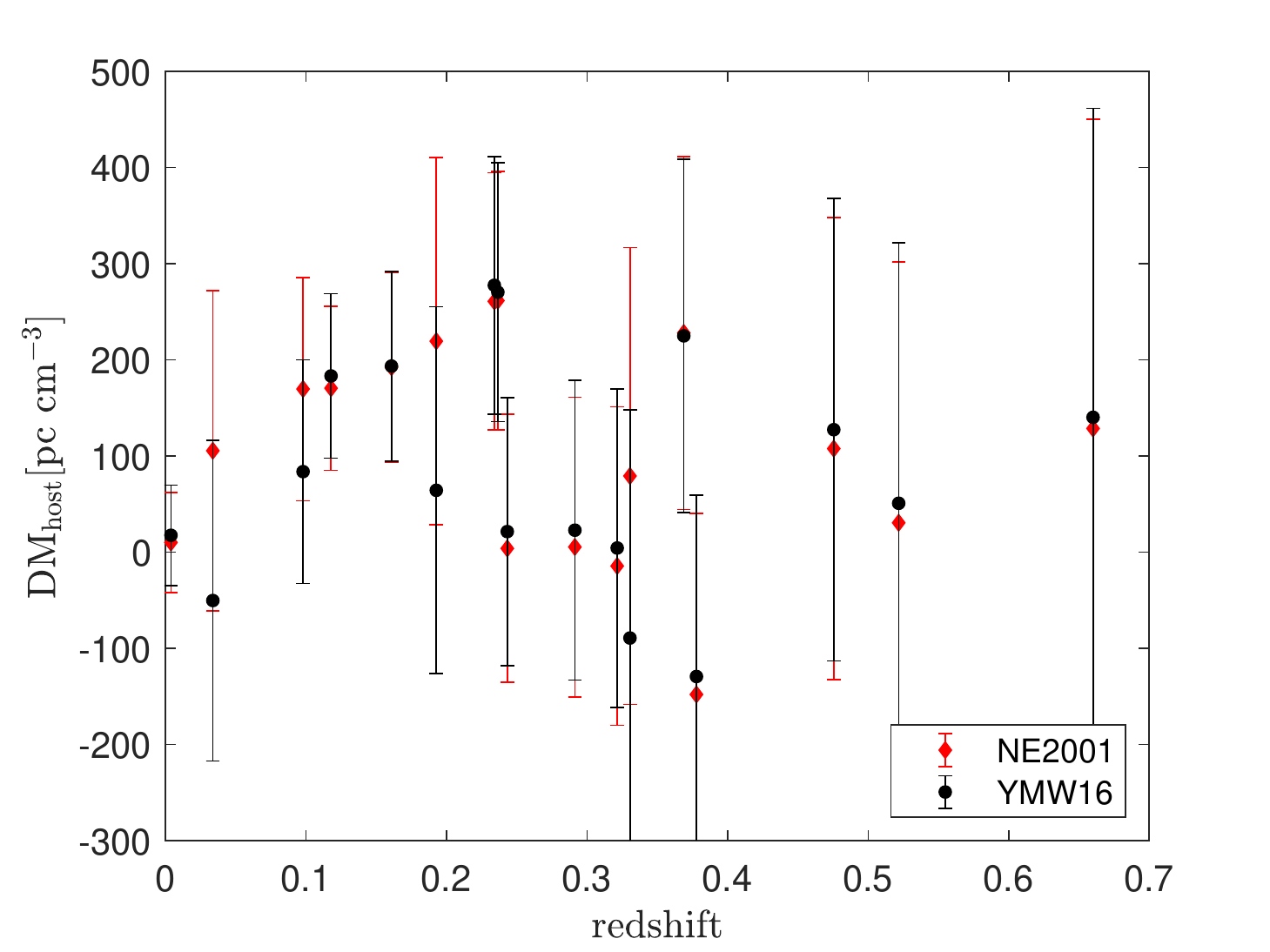}
 \includegraphics[width=0.48\textwidth]{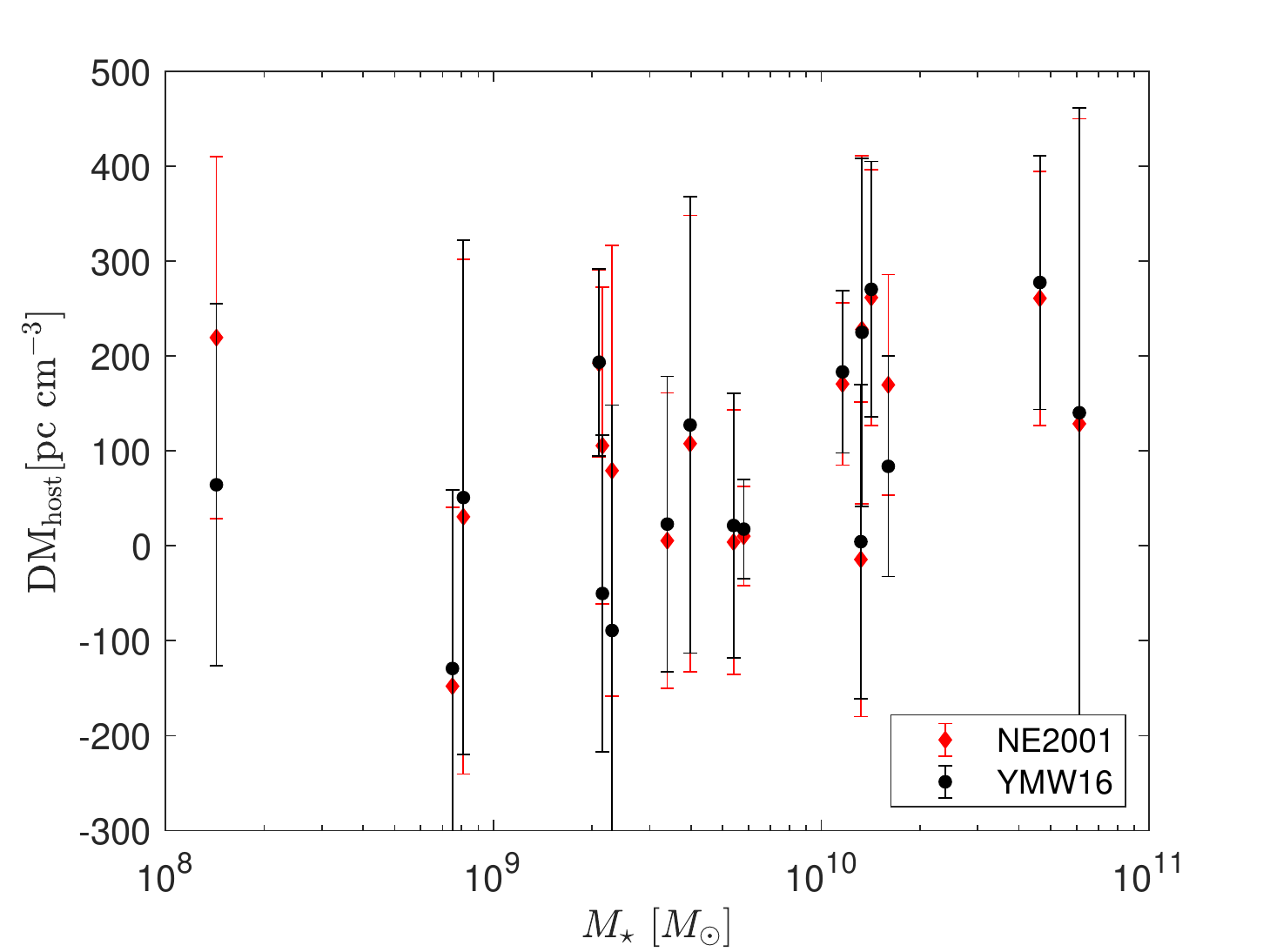}
 \includegraphics[width=0.48\textwidth]{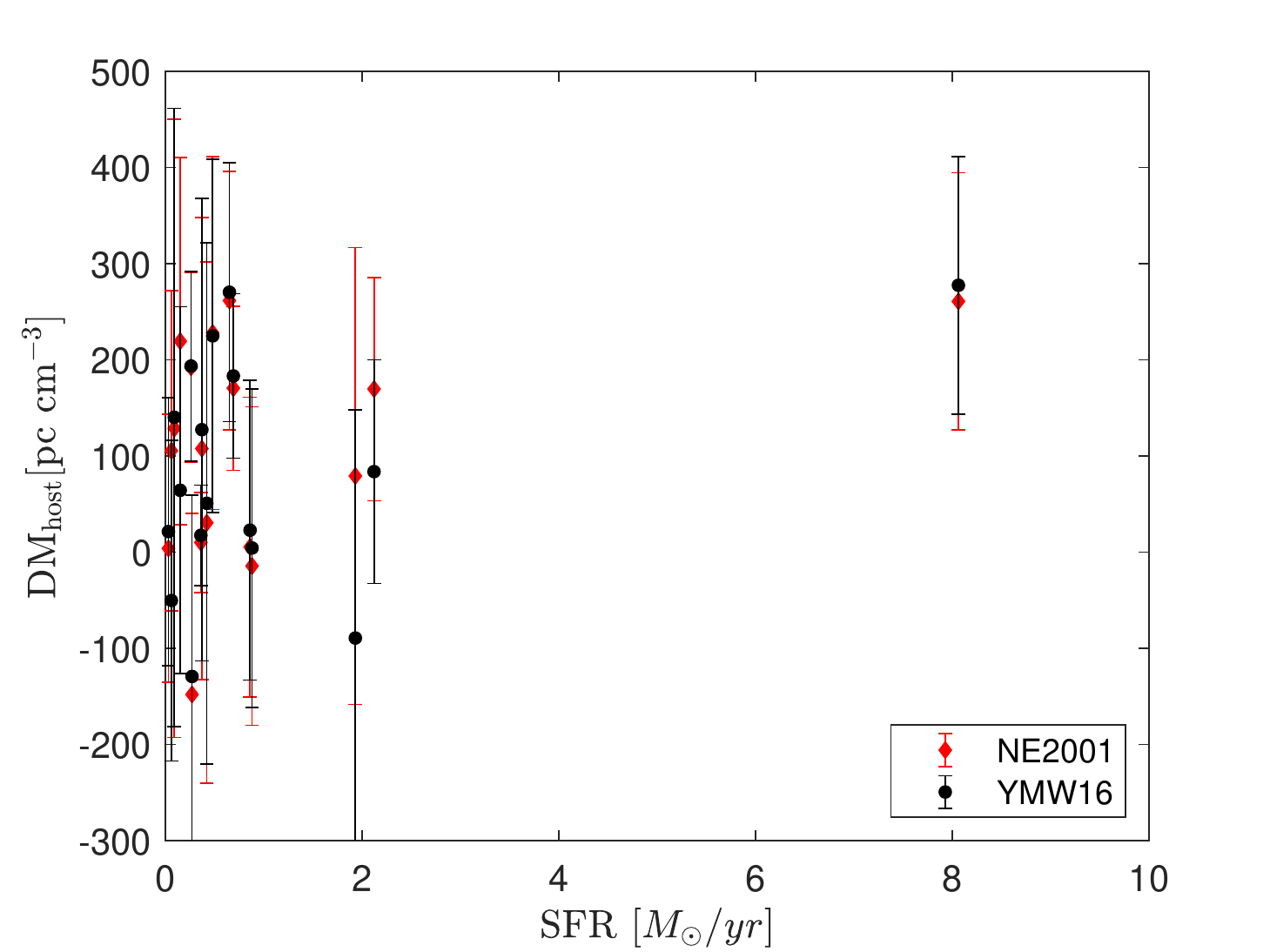}
 \includegraphics[width=0.48\textwidth]{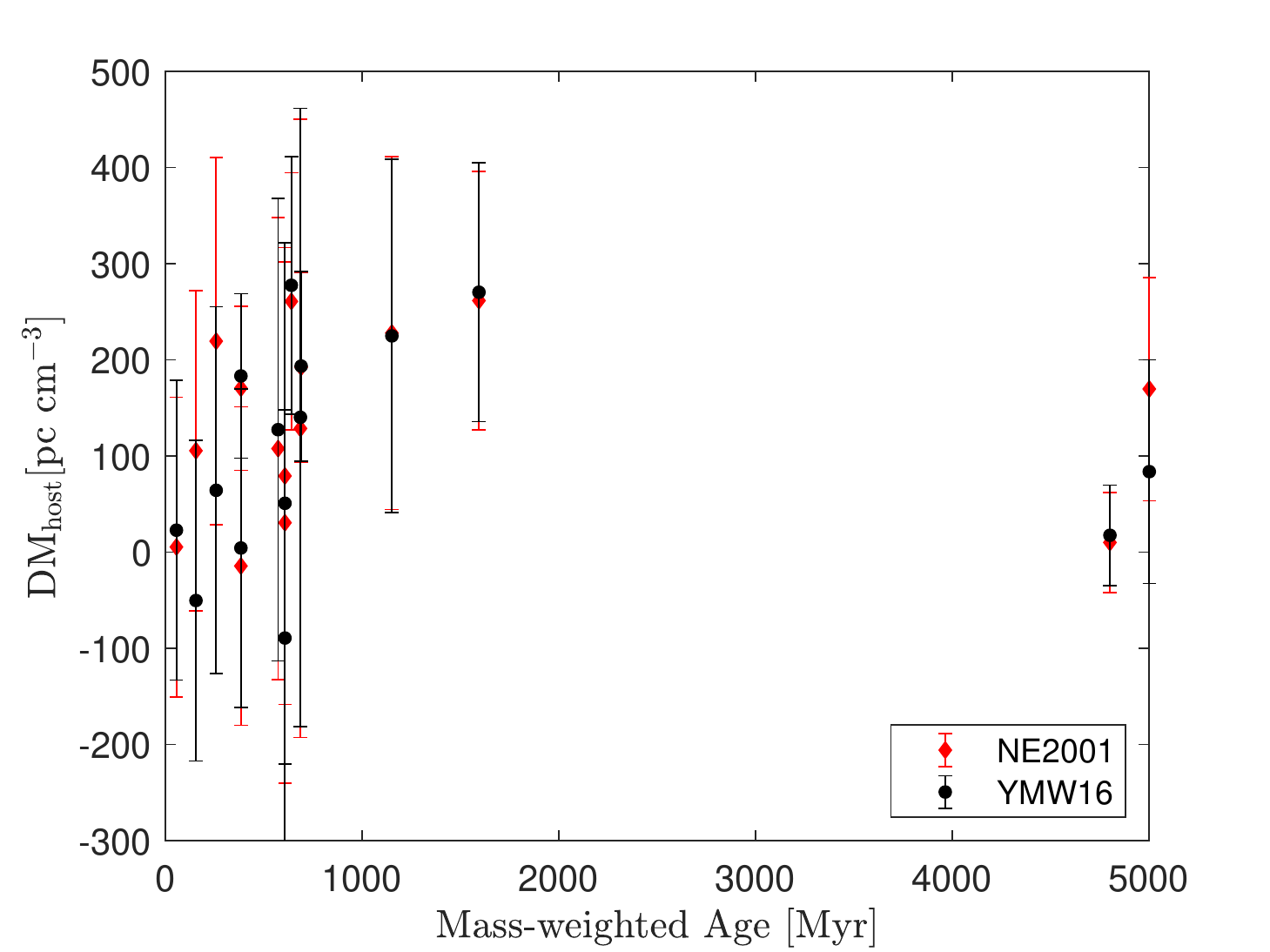}
 \includegraphics[width=0.48\textwidth]{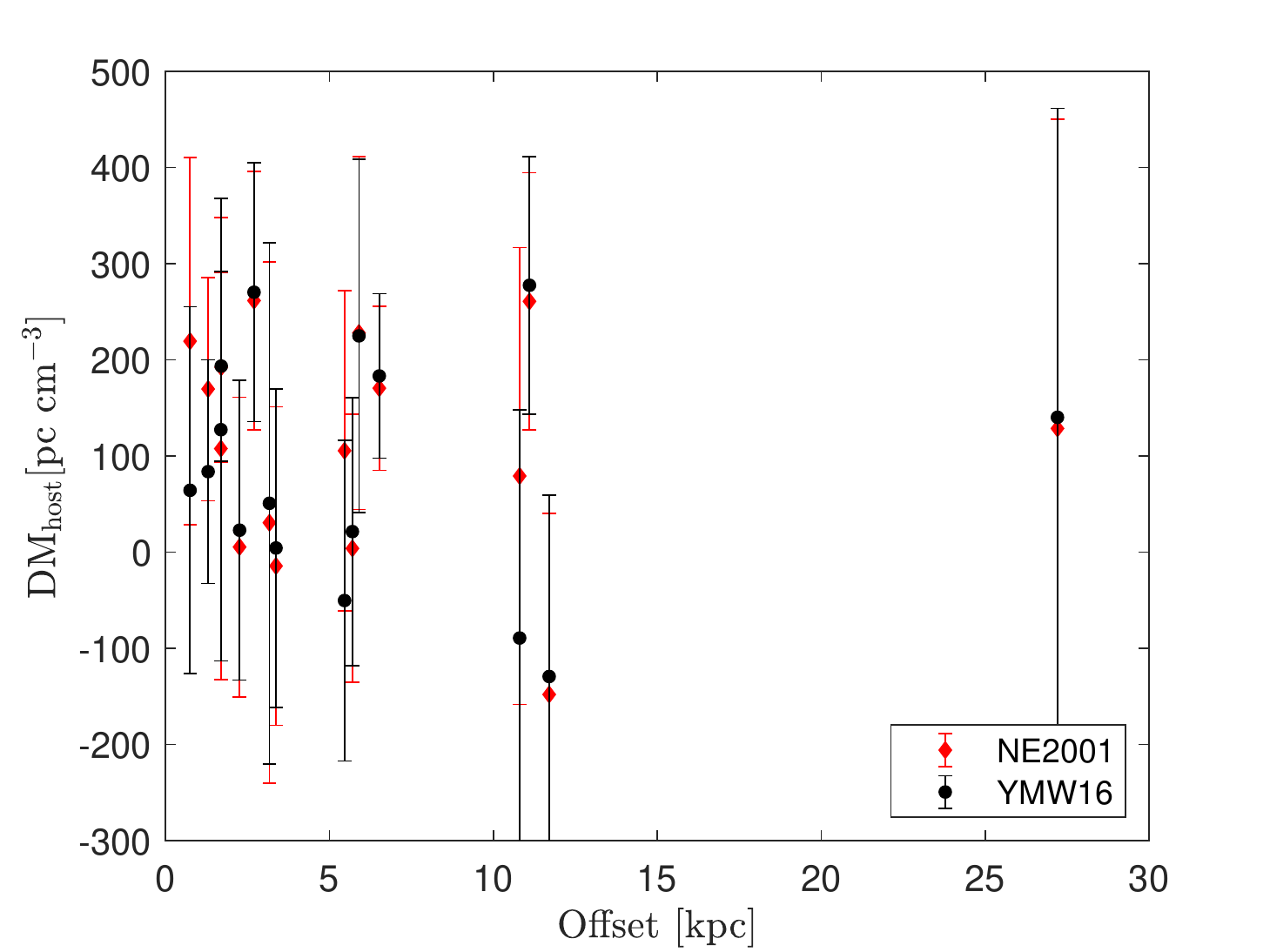}
 \includegraphics[width=0.48\textwidth]{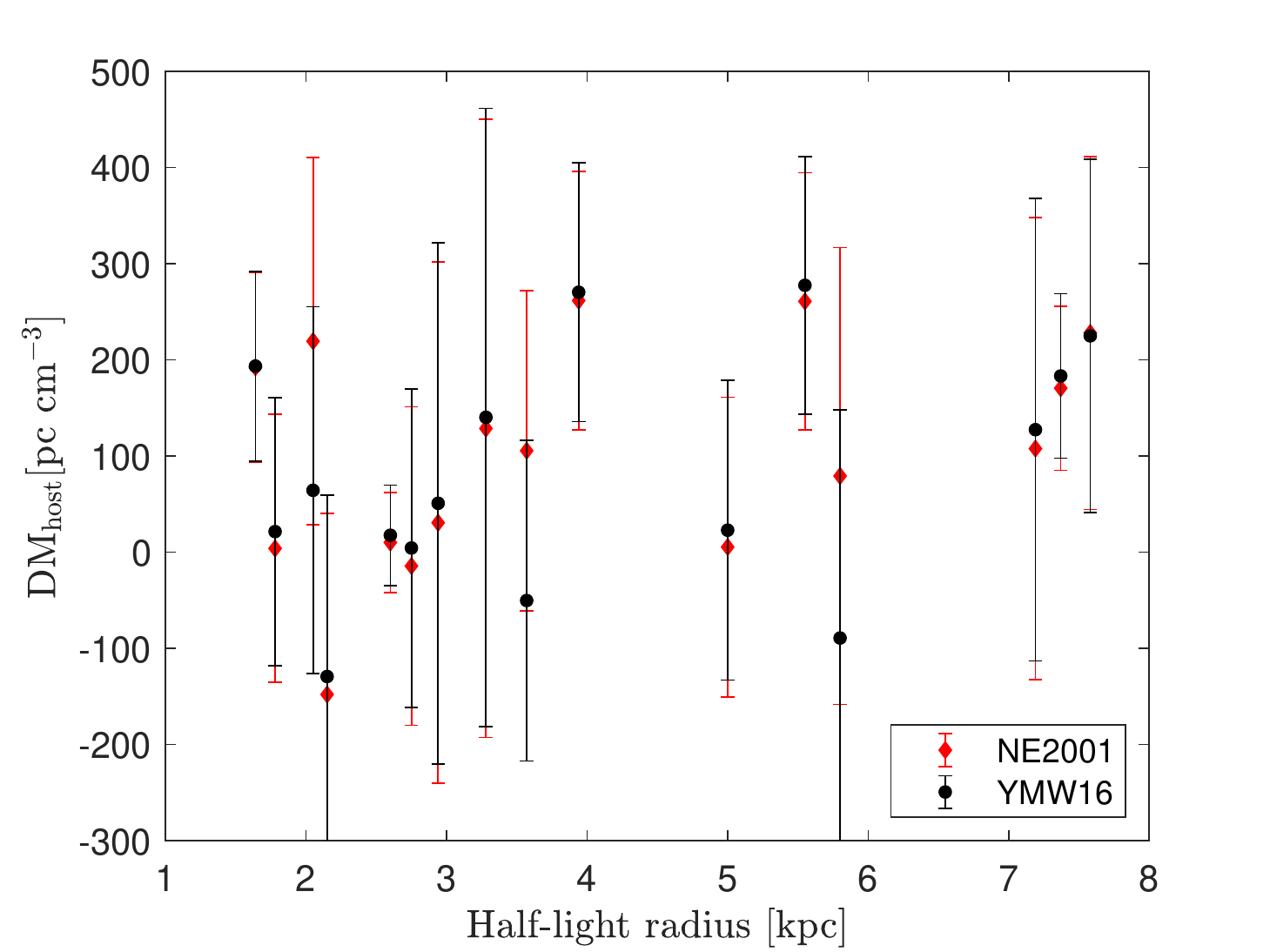}
 \caption{The correlations between ${\rm DM_{host}}$ and the properties of host galaxy. The DM of the Milky Way is calculated using two different electron models, i.e. the NE2001 model (red diamonds) and the YMW16 model (black circles).}\label{fig:DM_host}
\end{figure*}

\begin{table*}[htbp]
\centering
\caption{\small{The Spearman's correlation coefficients between ${\rm DM_{host}}$ and any parameter of the host galaxies.}}\label{tab:Spearman}
\begin{tabular}{ccccccc}
\hline\hline
& $z$ & $M_{\star}$ & SFR & Age & Offset & $R_{\rm eff}$  \\
\hline
NE2001& --0.23 & 0.36 & 0.18 & 0.37 & --0.17 & 0.37 \\
YMW16 & --0.04 & 0.54 & 0.21 & 0.43 & --0.09 & 0.19 \\
\hline
\end{tabular}
\end{table*}

In Figure \ref{fig:DM_host}, we plot the correlations between ${\rm DM_{host}}$ and the properties of host galaxies. In all subfigures, the vertical axes are ${\rm DM_{host}}$, and the horizonal axes are the redshift, the stellar mass, the SFR, the mass-weighted age, the offset from galactic center, and the half-light radius, respectively. In Table \ref{tab:Spearman}, we list the Spearman's correlation coefficients $\rho$ of six correlations \cite{Spearman:1904}. In general, $|\rho|<0.3$, $0.3<|\rho|<0.7$ and $|\rho|>0.7$ imply that the correlation is weak, moderate and strong, respectively \cite{Haldun:2018}. From Table \ref{tab:Spearman}, we see that there is no strong correlation between ${\rm DM_{host}}$ and any of the host galaxy parameters, neither in NE2001 model nor in YMW16 model. From the upper-left panel of Figure 2, we note that ${\rm DM_{host}}$ of the first 8 FRBs at $z<0.24$ is strongly linearly correlated with redshift. The Spearman's correlation coefficients are 1.0 and 0.8 for NE2001 model and YMW16 model, respectively. The positive ${\rm DM_{host}}-z$ correlation means that high-redshift FRBs generally have larger host DM than low-redshift FRBs, which may imply that high-redshift galaxies have richer diffuse gas than low-redshift galaxies. Due to the small FRB sample and large uncertainty, it's unclear if the ${\rm DM_{host}}-z$ correlation is intrinsic or not. For high-redshift FRBs ($z>0.24$), however, the correlation disappears. Therefore, we suspect that the linear ${\rm DM_{host}}-z$ correlation at $z<0.24$ may be happened by chance. From Table \ref{tab:Spearman}, we note that there is a moderate correlation between ${\rm DM_{host}}$ and the stellar mass of host galaxy. The positive ${\rm DM_{host}}-M_\star$ correlation implies that a more massive galaxy usually contributes a larger ${\rm DM_{host}}$. This is because more massive galaxies in general contain more diffuse gas. In addition, some other factors such as the age of host galaxy may also moderately affect ${\rm DM_{host}}$. A larger FRB sample is required to confirm or falsify the ${\rm DM_{host}}-M_\star$ correlation.

The central value of ${\rm DM_{host}}$ of FRB20190611B (galactic latitude $b=-33.6^{\circ}$, redshift $z=0.3778$) is somehow negative in both NE2001 model and YMW16 model, which implies that ${\rm DM_{MW,ISM}}$ and/or ${\rm DM_{IGM}}$ for this burst are/is overestimated. Since the ${\rm DM_{MW,ISM}}$ values of FRB 190611 calculated from both models are consistent with that of other FRBs located at the similar directions (such as FRB20190102C and FRB 20190711A), the most possibility is that ${\rm DM_{MW,ISM}}$ is accurate while ${\rm DM_{IGM}}$ is overestimated. The overestimation of ${\rm DM_{IGM}}$ may be caused by matter fluctuation. For FRB20180301A ($b=-5.8^{\circ}$, $z=0.3305$) and FRB20180916B ($b=4.0^{\circ}$, $z=0.0337$), the central values of ${\rm DM_{host}}$ calculated from YMW16 model are negative. This is because the YMW16 model may overestimate the Milky Way ISM contribution at low Galactic latitude \cite{KochOcker:2021fia}. Excluding the unphysical negative values, the mean and standard deviation of ${\rm DM_{host}}$ are $({\rm \overline{DM}_{host}},\sigma_{\rm DM_{host}})=(131.6, 92.0)$ ${\rm pc~cm}^{-3}$ for NE2001 model, and $({\rm \overline{DM}_{host}},\sigma_{\rm DM_{host}})=(120.1, 96.3)$ ${\rm pc~cm}^{-3}$ for YMW16 model.

\section{Constraints on the fraction of baryon mass}\label{sec:constrain}

The DM of IGM in equation (\ref{eq:DM_IGM}) contains the information of cosmology, thus can be used to study the Universe. In this section we use the well-localized FRBs to constrain the fraction of baryon mass in IGM, i.e. the parameter $f_{\rm IGM}(z)$. To test if $f_{\rm IGM}$ is redshift-dependent or not, we follow Li et al. \cite{Li:2019klc} and parameterize it as a slowly evolving function of redshift,
\begin{equation}
  f_{\rm IGM}=f_{\rm IGM,0}\left(1+\frac{\alpha z}{1+z}\right),
\end{equation}
where $f_{\rm IGM,0}$ is the fraction of baryon mass in the IGM at present day, and $\alpha$ is a constant.

In the previous section, we have shown that there is no strong correlation between ${\rm DM_{host}}$ and any of the host galaxy parameters. Therefore, there is no reason to parameterize ${\rm DM_{host}}$ as a function of one or some of the host galaxy parameters. The simplest and most straightforward assumption is that ${\rm DM_{host}}$ is a constant. We introduce an uncertainty term $\sigma_{\rm DM_{host}}$ to account for the possible deviation from the constant. The value of $\sigma_{\rm DM_{host}}$ is fixed to be the standard deviation of ${\rm DM_{host}}$ obtained in the previous section, i.e. $\sigma_{\rm DM_{host}}=92.0$ and 96.3 ${\rm pc~cm}^{-3}$ in NE2001 model and YMW16 model, respectively. This choice of $\sigma_{\rm DM_{host}}$, rather than the value calculated from equation (\ref{eq:sigma_host}), avoids double bias caused by the large uncertainty of ${\rm DM_{IGM}}$.

By fitting the observed DM to the theoretical prediction, the cosmological parameters can be constrained. The likelihood function is given by
\begin{equation}
  \mathcal{L}({\rm Data}|{\bm{\theta}})=\prod_{i=1}^N\frac{1}{\sqrt{2\pi}\sigma_{\rm total}}\exp\left(-\frac{1}{2}\chi^2\right),
\end{equation}
where $\bm{\theta}$ is the set of free parameters and `Data' represents the FRB sample, and
\begin{equation}
  \chi^2=\frac{[{\rm DM_{obs}}-{\rm DM_{MW}}-{\rm DM_{IGM}}-{\rm DM_{host}/(1+z)}]^2}{\sigma_{\rm total}^2},
\end{equation}
where ${\rm DM_{IGM}}$ is calculated from equation (\ref{eq:DM_IGM}), and the total uncertainty is given by \cite{Li:2019klc}
\begin{equation}\label{eq:sigma_total}
  \sigma_{\rm total}=\sqrt{\sigma_{\rm obs}^2+\sigma_{\rm MW}^2+\sigma_{\rm IGM}^2+\sigma_{\rm DM_{host}}^2/(1+z)^2}.
\end{equation}
The posterior probability density functions (PDFs) of the parameters are given by
\begin{equation}
  P({\bm{\theta}}|{\rm Data})\propto \mathcal{L}({\rm Data}|{\bm{\theta}})P_0({\bm\theta}),
\end{equation}
where $P_0({\bm\theta})$ is the prior of the parameters.

We calculate the posterior PDFs of the parameters using the publicly available python package \textsf{emcee}\footnote{https://emcee.readthedocs.io/en/stable/} \cite{Foreman-Mackey:2012any}. Note that $f_{\rm IGM,0}$ is completely degenerated with the Hubble constant $H_0$ and the baryon density $\Omega_{b}$, hence we fix the latter two parameters to the Planck 2018 values, i.e. $H_0=67.4~{\rm km~s^{-1}~Mpc^{-1}}$ and $\Omega_{b}=0.0493$ \cite{Aghanim:2018eyx}. In addition, $\Omega_m$ and $\Omega_\Lambda$ depict the background Universe and they have been tightly constrained by the Planck data. Therefore, we also fix them to the Planck 2018 values, namely, $\Omega_m=0.315$ and $\Omega_\Lambda=0.685$ \cite{Aghanim:2018eyx}. This leaves behind three free parameters $(f_{\rm IGM,0},\alpha,{\rm DM_{host}})$. We use a flat prior on all the free parameters: $f_{\rm IGM,0}\in \mathcal{U}(0,1)$, $\alpha\in \mathcal{U}(-2,2)$ and ${\rm DM_{host}}\in \mathcal{U}(0,300)~{\rm pc~cm^{-3}}$.

\begin{figure*}[htbp]
 \centering
 \includegraphics[width=0.48\textwidth]{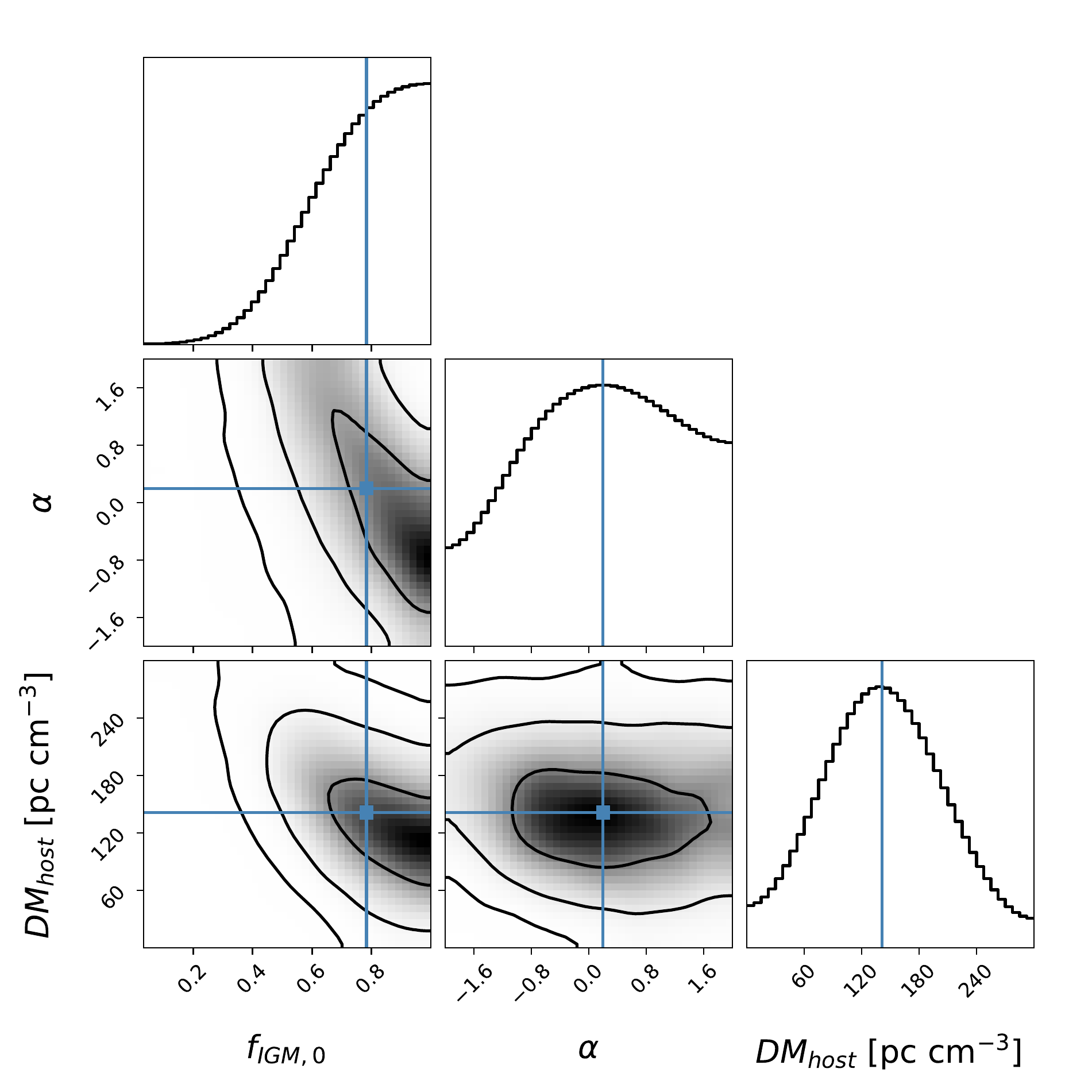}
 \includegraphics[width=0.48\textwidth]{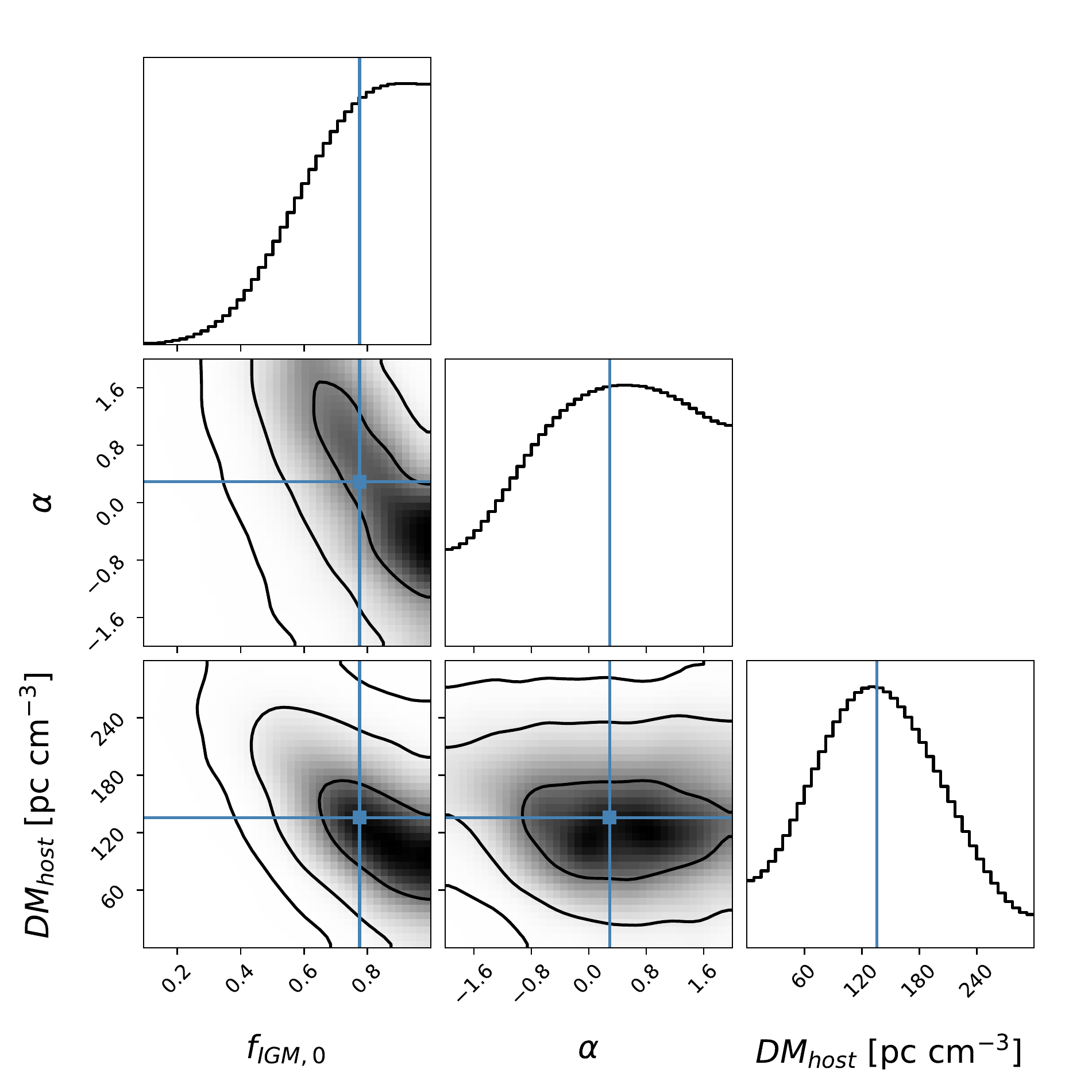}
 \caption{The posterior PDFs and confidence contours on the free parameters ($f_{\rm IGM,0}, \alpha, {\rm DM_{host}}$) by assuming a constant ${\rm DM_{host}}$. Left panel: NE2001 model; right panel: YMW16 model.}\label{fig:constrain1}
\end{figure*}

The best-fitting parameters ($f_{\rm IGM,0}, \alpha, {\rm DM_{host}}$) are listed in Table \ref{tab:parameters1}, and the marginalized posterior PDFs and 2-dimensional marginalized confidence contours of the parameter space are plotted in Figure \ref{fig:constrain1}. FRBs with negative ${\rm DM_{host}}$ values are excluded in the fitting. For NE2001 model, we obtain $f_{\rm IGM,0}=0.78_{-0.19}^{+0.15}$, $\alpha=0.20_{-1.14}^{+1.15}$ and ${\rm DM_{host}}=141.3_{-55.8}^{+59.8}~{\rm pc~cm^{-3}}$, where the uncertainties are given at $1\sigma$ confidence level. For YMW16 model, we obtain $f_{\rm IGM,0}=0.78_{-0.19}^{+0.15}$, $\alpha=0.29_{-1.18}^{+1.10}$ and ${\rm DM_{host}}=135.8_{-60.4}^{+65.6}~{\rm pc~cm^{-3}}$. In both models, $f_{\rm IGM,0}$ and ${\rm DM_{host}}$ can be tightly constrained. Although the constraint on $\alpha$ is loose, the best-fitting $\alpha$ prefers a positive value, which is consistent with the requirement that $f_{\rm IGM}$ mildly increases with redshift \cite{McQuinn:2013tmc,Li:2019klc}. The two Milky Way electron models give very consistent results within $1\sigma$ uncertainty, which is because high-latitude FRBs have much larger weights than low-latitude FRBs in the fitting, while both models give consistent ${\rm DM_{MW,ISM}}$ values for high-latitude FRBs. Based on the limited number of FRBs and large uncertainty, there is no evidence for the redshift evolution of baryon mass fraction in IGM.

\begin{table*}[htbp]
\centering
\caption{The best-fitting parameters ($f_{\rm IGM,0}, \alpha, {\rm DM_{host}}$) by assuming a constant ${\rm DM_{host}}$.}\label{tab:parameters1}
\arrayrulewidth=0.5pt
\renewcommand{\arraystretch}{1.2}
{\begin{tabular}{cccc} 
\toprule
& $f_{\rm IGM,0}$ & $\alpha$ & ${\rm DM_{host}~[pc~cm^{-3}]}$ \\
\hline
NE2001& $0.78_{-0.19}^{+0.15}$ & $0.20_{-1.14}^{+1.15}$ & $141.3_{-55.8}^{+59.8}$ \\
YMW16 & $0.78_{-0.19}^{+0.15}$ & $0.29_{-1.18}^{+1.10}$ & $135.8_{-60.4}^{+65.6}$ \\
\bottomrule
\end{tabular}}
\end{table*}

\begin{figure*}[htbp]
 \centering
 \includegraphics[width=0.48\textwidth]{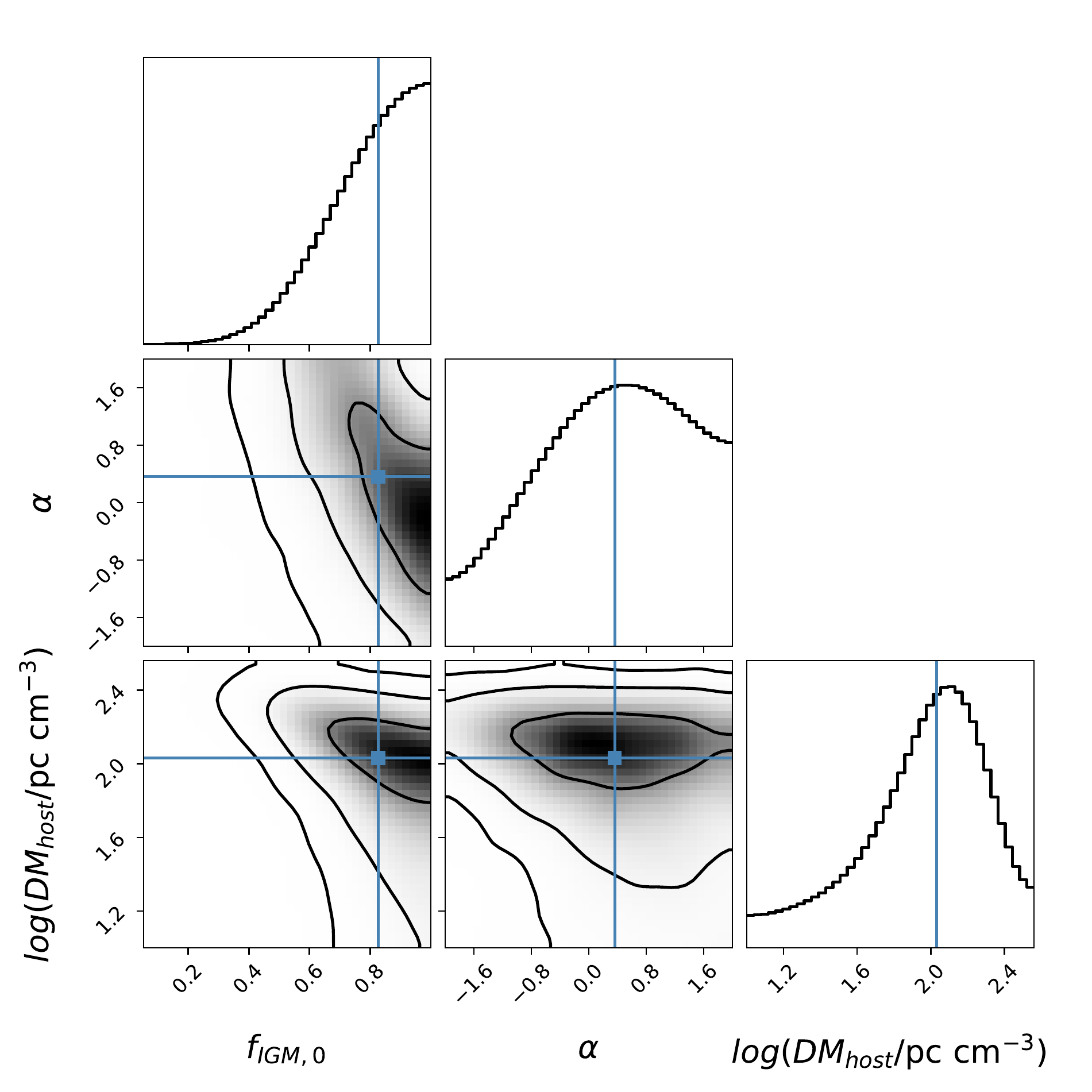}
 \includegraphics[width=0.48\textwidth]{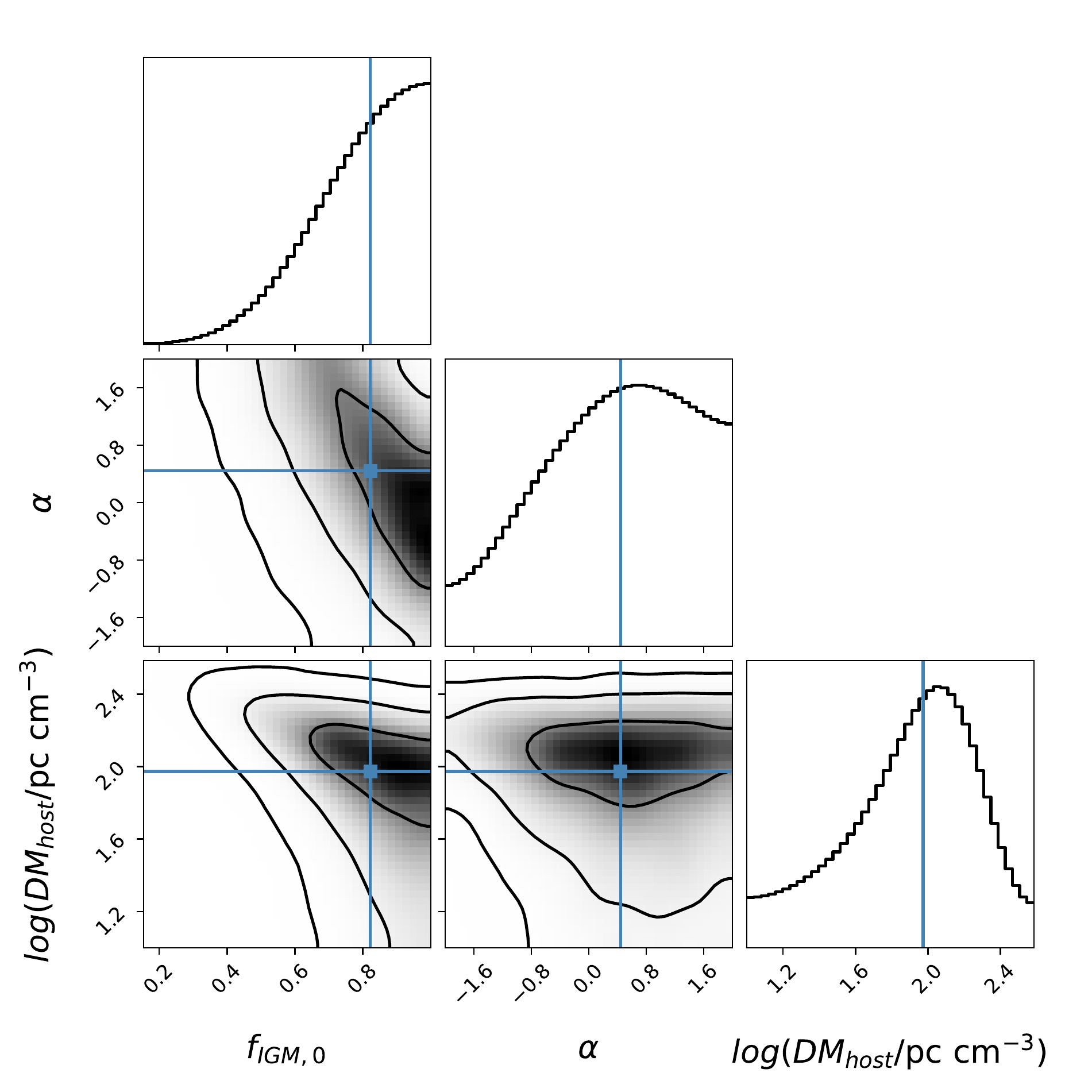}
 \caption{The posterior PDFs and confidence contours on the free parameters ($f_{\rm IGM,0}, \alpha, {\rm DM_{host}}$) by assuming a log-normal distribution on ${\rm DM_{host}}$. Left panel: NE2001 model; right panel: YMW16 model.}\label{fig:constrain2}
\end{figure*}

\begin{figure*}[htbp]
 \centering
 \includegraphics[width=0.48\textwidth]{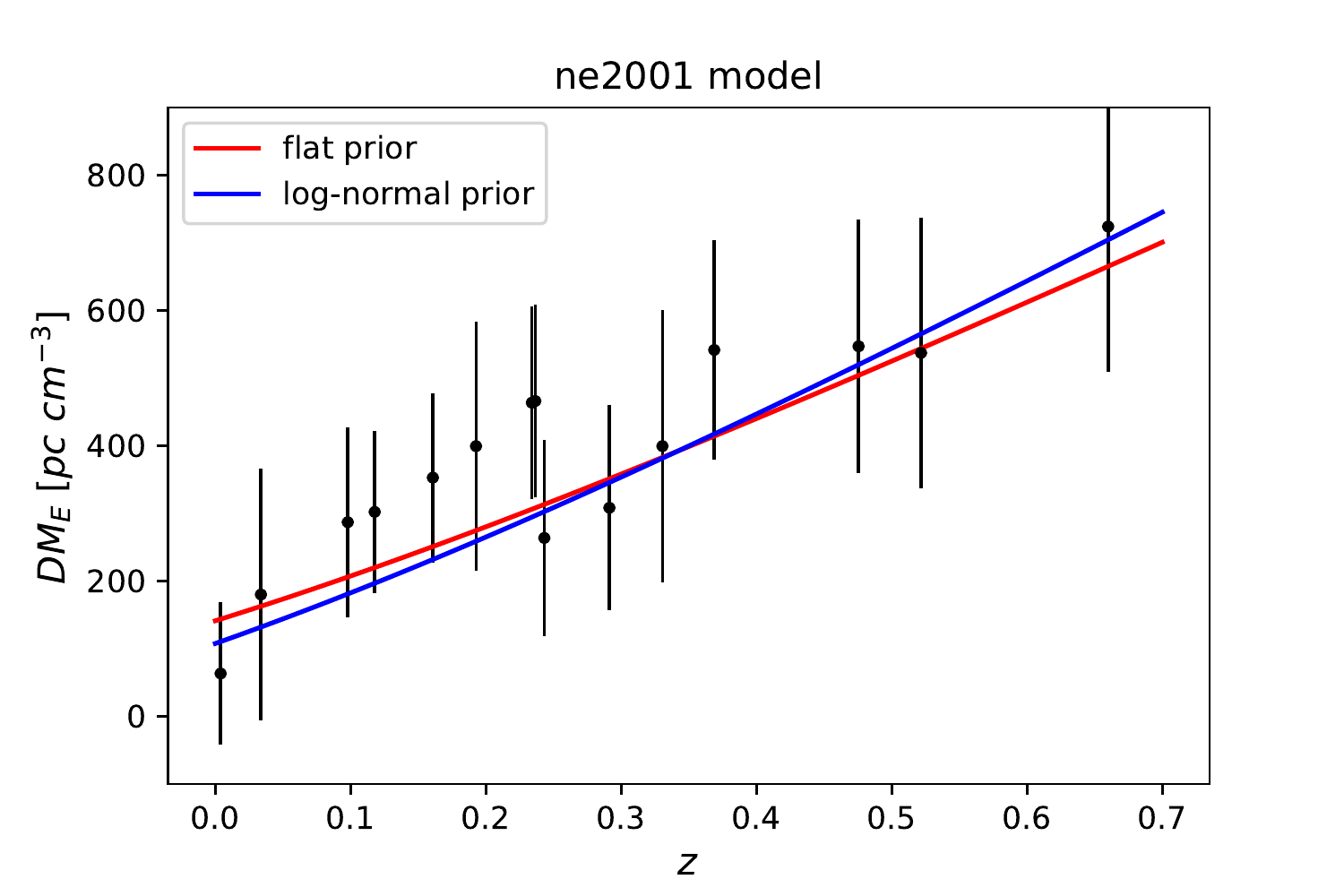}
 \includegraphics[width=0.48\textwidth]{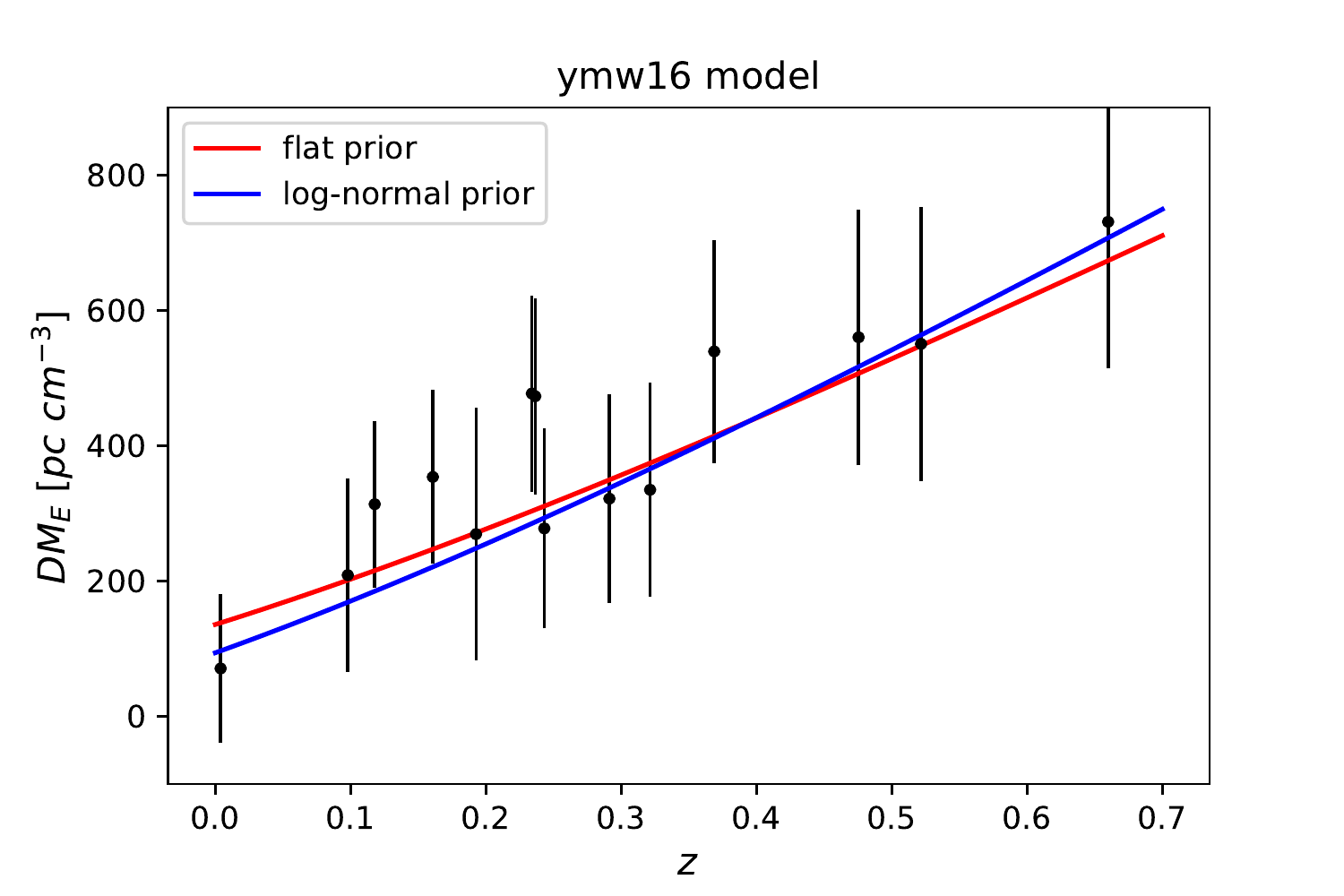}
 \caption{The best-fitting curves to the extragalactic DM for NE2001 model (left) and YMW16 model (right). Black dots with $1\sigma$ error bars are the data points, red and blue lines are the best-fitting curves assuming flat prior and log-normal prior on ${\rm DM_{host}}$, respectively.}\label{fig:fitting}
\end{figure*}

In fact, ${\rm DM_{host}}$ can vary significantly from burst to burst. Hence it is not a good approximation to assume a constant ${\rm DM_{host}}$. A more reasonable way to deal with ${\rm DM_{host}}$ is to model it as a probability distribution and marginalize over the free parameters. It is shown that ${\rm DM_{host}}$ can be fitted with log-normal distribution from theory and cosmological simulations \cite{Macquart:2020lln,Jaroszynski:2020kqy,Zhang:2020mgq}. Therefore, instead of assuming a constant ${\rm DM_{host}}$, we model it as a log-normal distribution centered at ${\rm DM_{host}}$ and refit the data. This is equivalent to use a log-normal prior on the parameter ${\rm DM_{host}}$ in the MCMC fitting. The best-fitting results are listed in Table \ref{tab:parameters2}. The marginalized posterior PDFs and 2-dimensional marginalized confidence contours of the parameter space are plotted in Figure \ref{fig:constrain2}. The best-fitting curves to the extragalactic DM $({\rm DM_E\equiv DM_{obs}-DM_{MW}=DM_{IGM}+DM_{host}}/(1+z))$ for NE2001 model and YMW16 model are plotted in the left and right panels of Figure \ref{fig:fitting}, respectively. Using log-normal prior, we obtain a larger $f_{\rm IGM}$ value (0.83 vs. 0.78) and a smaller ${\rm DM_{host}}$ ($\sim 100$ vs. $\sim 140$ pc cm$^{-3}$) than using flat prior. Using log-normal prior, the best-fitting $f_{\rm IGM}$ is more consistent with the fiducial value (0.84). This confirms that assuming a log-normal distribution on ${\rm DM_{host}}$ is more reasonable than assuming a constant value.

\begin{table*}[htbp]
\centering
\caption{The best-fitting parameters ($f_{\rm IGM,0}, \alpha, {\rm DM_{host}}$) by assuming a log-normal distribution on ${\rm DM_{host}}$.}\label{tab:parameters2}
\arrayrulewidth=0.5pt
\renewcommand{\arraystretch}{1.2}
{\begin{tabular}{cccc} 
\toprule
& $f_{\rm IGM,0}$ & $\alpha$ & ${\rm DM_{host}~[pc~cm^{-3}]}$ \\
\hline
NE2001& $0.83_{-0.17}^{+0.12}$ & $0.36_{-1.13}^{+1.02}$ & $107.7_{-62.9}^{+65.3}$ \\
YMW16 & $0.83_{-0.17}^{+0.12}$ & $0.44_{-1.15}^{+1.00}$ & $94.0_{-59.1}^{+69.0}$ \\
\bottomrule
\end{tabular}}
\end{table*}

\section{Discussion and Conclusions}\label{sec:conclusions}

In this paper, we investigated the host galaxy DM using the well-localized FRBs. We tested the possible correlations between ${\rm DM_{host}}$ and six properties of host galaxies: the redshift, the stellar mass, the star-formation rate, the age of galaxy, the offset of FRB site from galactic center, and the half-light radius. We found that there is no strong correlation between ${\rm DM_{host}}$ and any of the parameters. Luo et al. \cite{Luo:2018tiy} pointed out that ${\rm DM_{host}}$ is proportional to the square-root of the SFR of host galaxy. However, we found no correlation between them in the 17 well-localized FRBs. The host of FRB20191001A is an active galaxy, with ${\rm SFR}=8.06\pm 2.42~M_\odot~{\rm yr}^{-1}$. For the rest 16 FRBs, the SFRs of host galaxies are small (${\rm SFR}\lesssim 2~M_\odot~{\rm yr}^{-1}$). There is a big gap in the SFR range $2\sim 8~M_\odot~{\rm yr}^{-1}$. We can't exclude the possible existence of correlation between ${\rm DM_{host}}$ and SFR if FRB sample is enlarged in the future.

The FRB sample contains host galaxies of very different properties. For example, the galaxy types vary from burst to burst, the stellar masses of host galaxies span a wide range from $10^8M_\odot$ to $10^{11}M_\odot$, and the galaxy ages range from decades Myr to thousands Myr. Since the available FRB sample is very small, we have to combine all the FRBs together to study the correlations. When studying the correlation between host DM and other properties (such as ${\rm DM_{host}}-z$ correlation), it is better to choose FRBs whose other properties (such as galaxy type, stellar mass, SFR, etc.) are similar. Only in this way we can make a fair comparison. However, due to the small FRB sample, we can't do this at present. An alternative way is to study the multi-dimensional correlations. But this also requires a large FRB sample. Thus multi-dimensional correlations are not considered here. If the host galaxy is a spiral galaxy, one of the main factors that may affect ${\rm DM_{host}}$ is the inclination angle. A edge-on galaxy is expected to contribute a larger value of ${\rm DM_{host}}$ than a face-on galaxy. Unfortunately, except for FRB20190608B \cite{2020arXiv200513158C}, the rest FRBs have no observation on inclination angle.

The ${\rm DM_{host}}$ values obtained by subtracting the contributions of MW and IGM from the observed DM have large uncertainty. The uncertainty on ${\rm DM_{host}}$ is dominated by the uncertainty on ${\rm DM_{IGM}}$. For low-latitude FRBs, the ${\rm DM_{MW,ISM}}$ term also introduces a large uncertainty. Because both Milky Way electron models do not provide the uncertainty of ${\rm DM_{MW,ISM}}$ directly, we just simply adopt the difference of ${\rm DM_{MW,ISM}}$ values calculated from two electron models as $\sigma_{\rm MW,ISM}$. This is reasonable for high-latitude FRBs, since both electron density models give consistent ${\rm DM_{MW,ISM}}$ values. For low-latitude FRBs, however, YMW16 model gives a much higher value of ${\rm DM_{MW,ISM}}$ than NE2001 model. Koch Ocker et al. \cite{KochOcker:2021fia} pointed out that YMW16 model may overestimate ${\rm DM_{MW,ISM}}$ at low latitude. With more Galactic plane pulsars discovered by e.g. the FAST telescope \cite{Han:2021ekd}, ${\rm DM_{MW,ISM}}$ at low latitude is expected to be modeled more accurately in the future. In addition, FRB source may also contribute a non negligible DM value, thus introduces additional uncertainty. However, without independent observations, the FRB source contribution is indistinguishable from host galaxy contribution. Therefore, we don't distinguish them and treat them as a whole. Unless we can observe the DM of FRB source directly in the future, this part can be separated and the uncertainty can be reduced. Of course, if we can observe the host DM directly (through e.g. optical/UV observations) \cite{2020arXiv200513158C}, the uncertainty of ${\rm DM_{host}}$ can be further reduced.

We assumed a constant value of ${\rm DM_{host}}$ for the FRBs, and used it to constrain the fraction of baryon mass in IGM. We found no strong evidence for the redshift dependence of $f_{\rm IGM}$, and obtained a consistent constraint in both Milky Way electron models, i.e. $f_{\rm IGM,0}=0.78_{-0.19}^{+0.15}$. Our results are consistent with that of Ref.\cite{Li:2020qei}, which used a small sample of localized FRBs to constrain the fraction of baryon mass in IGM, and obtained $f_{\rm IGM,0}=0.84_{-0.22}^{+0.16}$ from five FRBs, and $f_{\rm IGM,0}=0.74_{-0.18}^{+0.24}$ from three non-repeating FRBs. The central value we obtained here ($f_{\rm IGM,0}=0.78$) is somewhat smaller than some previous observations, e.g. $f_{\rm IGM,0}\approx 0.83$ \cite{Fukugita:1997bi}, but they are still consistent within $1\sigma$ uncertainty. One reason why we obtain a smaller $f_{\rm IGM,0}$ value may be that the posterior probability density function of $f_{\rm IGM,0}$ is non-symmetric (see Figure \ref{fig:constrain1}). From the posterior probability density function we can see that the probability of $f_{\rm IGM,0}$ being smaller than 0.78 is suppressed. The best fitting values of ${\rm DM_{host}}$ we obtained are $141.3_{-55.8}^{+59.8}$ and $135.8_{-60.4}^{+65.6}~{\rm pc~cm^{-3}}$ for NE2001 and YMW16 models, respectively. These values are consistent with Ref.\cite{Macquart:2020lln}, which obtained a range of ${\rm DM_{host}}\in (20,200) {\rm pc~cm^{-3}}$ for non-repeating FRBs. They are also consistent with the host DM of FRB20190608B obtained from optical/UV observations, i.e. $137\pm 43~{\rm pc~cm^{-3}}$ \cite{2020arXiv200513158C}. If we model ${\rm DM_{host}}$ as a log-normal distribution instead of a constant, we obtained $f_{\rm IGM,0}=0.83_{-0.17}^{+0.12}$, which is well consistent with the fiducial value. In this case, we obtained a smaller average ${\rm DM_{host}}$ value in both Milky Way electron models, i.e. ${\rm DM_{host}}\sim 100~{\rm pc~cm^{-3}}$.

From equation \ref{eq:DM_IGM}, we know that $f_{\rm IGM}$ and $H_0$ are completely degenerated, namely $f_{\rm IGM}$ is inversely proportional to $H_0$. Therefore, a larger $H_0$ value will lead to a smaller $f_{\rm IGM}$ value. The $H_0$ value measured from local data ($H_0=73.48~{\rm km~s^{-1}~Mpc^{-1}}$) \cite{Riess:2018uxu} is in tension with the Planck value ($H_0=73.48~{\rm km~s^{-1}~Mpc^{-1}}$) \cite{Aghanim:2018eyx} at more than $3\sigma$. If we use the local $H_0$ value, we obtain $f_{\rm IGM,0}=0.76_{-0.16}^{+0.11}$, compared with $f_{\rm IGM,0}=0.83_{-0.17}^{+0.12}$ if the Planck value is used. We see that $f_{\rm IGM,0}$ is sensitive to the $H_0$ value. A biased $H_0$ will lead to a biased estimation of $f_{\rm IGM,0}$. Therefore, an accurate $H_0$ value is required in order to accurately constrain $f_{\rm IGM,0}$.



\end{multicols}

\vspace{-1mm}
\centerline{\rule{80mm}{0.5pt}}
\vspace{2mm}

\begin{multicols}{2}

\bibliographystyle{cpc-hepnp-1}
\bibliography{reference}

\end{multicols}

\end{document}